\documentclass[a4paper,11pt]{article}
\usepackage{pos}
\usepackage{psfrag}

\newcommand{\al}{\alpha}
\newcommand{\as}{\alpha_{\mathrm{s}}}
\newcommand{\asb}{\bar\alpha_{\mathrm{s}}}

\newcommand{\de}{\partial}
\newcommand{\dif}{\mathrm{d}}

\newcommand{\GeV}{\text{GeV}}
\newcommand{\ga}{\gamma}
\newcommand{\half}{{\textstyle\frac{1}{2}}}
\renewcommand{\hom}{{\textstyle\frac{\omega}{2}}}
\newcommand{\imf}{\phi}                  

\newcommand{\kt}{{\boldsymbol k}}
\newcommand{\lra}{\leftrightarrow}

\newcommand{\om}{\omega}
\newcommand{\ord}[1]{\mathcal{O}\left(#1\right)}

\newcommand{\ui}{\mathrm{i}}             

\newcommand{\ggfb}{G}
\newcommand{\ggf}{\mathcal{G}}
\newcommand{\krn}{\mathcal{K}}

\title{Small $x$ resummation of photon impact factors and the
  $\gamma^* \gamma^*$ high energy scattering}

\ShortTitle{RGI photon impact factors}

\author*[a]{Dimitri Colferai}
\author[b]{Anna Sta\'sto}
\author[b]{Wanchen Li}

\affiliation[a]{Department of Physics, University of Florence and INFN Florence,\\
  Via Sansone 1, Sesto Fiorentino, Italy}
\affiliation[b]{Department of Physics, Penn State University,\\
      University Park, PA 16802, USA}

\emailAdd{dimitri.colferai@unifi.it}
\emailAdd{ams52@psu.edu}
\emailAdd{wul161@psu.edu}

\abstract{I present the renormalization group improved collinear
  resummation of the photon-gluon impact factors.
We construct the resummed cross section for virtual photon-photon
($\gamma^*\gamma^*$) scattering which incorporates the impact factors
and BFKL gluon Green's function up to the next-to-leading logarithmic
accuracy in energy.
The impact factors include important kinematical effects which are
responsible for the most singular poles in Mellin space at
next-to-leading order. Further conditions on the resummed cross
section are obtained by requiring the consistency with the collinear
limits. Our analysis is consistent with previous impact factor
calculations at NLO, apart from a new term proportional to $C_F$ that
we find for the longitudinal polarization. Finally, we use the
resummed cross section to compare with the LEP data on the
$\gamma^*\gamma^*$ cross section and with previous calculations. The
resummed result is lower than the leading logarithmic approximation
but higher than the pure next-to-leading one, and is consistent with
the experimental data. 
}

\FullConference{31st International Workshop on Deep Inelastic Scattering (DIS2024)\\
 8–12 April 2024\\
Grenoble, France\\}

\begin{document}
\maketitle

\section{Introduction}

In this talk I present the calculation of the renormalization-group
improved (RGI) photon impact factors within the high-energy factorization
approach. Such impact factors, together with the BFKL gluon Green
function (GGF), enable us to compute the virtual photon cross section
in the high energy limit, resumming to all perturbative orders the leading
$(\as\log s)^n$ and next-to-leading $\as(\as\log s)^n$ logarithms of
the energy.

The BFKL expansion~\cite{BFKL} is known to be affected by large and negative
corrections at next-to-leading order, and it is known since long that
the main cause of the size of such corrections is due to large collinear
terms that need to be resummed as well~\cite{rgiggf}.

In momentum space, the high-energy factorization formula reads
\begin{equation}\label{sigmaMomSp}
  \sigma^{(jk)}(s,Q_1,Q_2) = \int\dif^2\kt\;\dif^2\kt' \;
    \imf^{(j)}(Q_1,\kt)\, \ggfb(s,\kt,\kt_0)\, \imf^{(k)}(Q_2,\kt_0) \;,
\end{equation}
where $j,k\in\{L,T\}$ denote the polarizations of the two photons,
$q_1,q_2$ their momenta and $Q_i^2\equiv -q_i^2>0:i=1,2$ their
virtualities.

The gluon Green's function $\ggfb(s,\kt,\kt_0) $, which depends on the
transverse gluon momenta $\kt$ and $\kt_0$ and energy squared
$s\equiv(q_1+q_2)^2$, satisfies the evolution equation that can be
written in the following form
\begin{equation}
  \frac{\partial }{\partial \log s} \ggfb(s,\kt,\kt_0) =  \int \dif^2
  \kt' \; \krn( \kt,\kt') \, \ggfb(s,\kt',\kt_0) \; ,
  \label{eq:BFKL_Green_mom}
\end{equation}
where the function $\krn$ is the BFKL kernel which has the following
perturbative expansion
\begin{equation}
\krn = \asb \krn_0 + \asb^2 \krn_1 + \dots \; .
    \label{eq:kernel_expansion}
\end{equation}
In the above equation, we introduced the rescaled strong coupling
\(\asb = \frac{\alpha_s N_c}{\pi} \) where $N_c$ is the number of
colors. In QCD the kernel is known at leading 
and next-to-leading 
order.  It is
customary to use the Mellin transform to obtain the kernel eigenvalue
\begin{equation}
\asb \chi(\gamma) =  \int \dif\kt^{\prime 2} \;
\bigg(\frac{\kt^{\prime2}}{\kt^2}\bigg)^{\gamma} \,  \krn(\kt,\kt') \,
, 
\label{eq:chi}
\end{equation}
with the corresponding perturbative expansion corresponding to
eq.~\eqref{eq:kernel_expansion}
\begin{equation}\label{asbchi}
     \chi(\gamma) =  \chi_0(\gamma) + \asb \chi_1(\gamma)+\dots \; .
\end{equation}
The leading  order kernel's eigenvalue reads
\begin{equation}\label{chi0}
  \chi_0(\gamma) =2\psi(1) - \psi(\gamma) - \psi(1-\gamma) \; ,
\end{equation}
where \(\psi(z) = \Gamma'(z)/\Gamma(z) \) is the polygamma function,
and $\psi(1)=-\gamma_E$. The behaviour of the kernel
$\krn_0(\kt,\kt')\sim 1/\kt^2$ for $\kt\gg\kt'$ reflects itself in the
presence of a simple pole of $\chi_0(\ga)\sim 1/\gamma$ for
$\gamma\to 0$. At higher orders, on the basis of collinear QCD
dynamics~\cite{DGLAP}, one expects that the higher-order
kernels behave like $\krn_n(\kt,\kt')\sim(1/\kt^2)\log^{n-1}(\kt/\kt')$
for $\kt\gg\kt'$ and thus
$\chi_{n}(\gamma)\sim 1/\gamma^{n+1}$. However, the actual
calculations yield $\chi_{n}(\gamma)\sim 1/\gamma^{2n+1}$.

The reason for such spurious poles has been found in the ``wrong''
choice of energy scale $s_0$ in the BFKL logarithms $\log(s/s_0)$.
In fact, if one rewrites the high-energy factorization formula in
Mellin space, where $\gamma$ is conjugated to virtualities and
transverse momenta, while $\omega$ is conjugated to the energy $s$,
one diagonalizes the convolutions into simple products:
  \begin{align}
  \sigma^{(jk)}(s,Q_1,Q_2) &= \frac1{2\pi Q_1 Q_2}
    \int\frac{\dif \omega}{2\pi\ui} \left(\frac{s}{s_0(p)}\right)^\om
    \int\frac{\dif \gamma}{2\pi\ui}
    \left(\frac{Q_1^2}{Q_2^2}\right)^{\gamma-\half}
    \frac{
      \imf^{(j)}(\gamma;p)\,\imf^{(k)}(1-\gamma;-p)}{\om-\asb\chi(\gamma;p)}
    \label{sigmaBFKL}
  \end{align}
In eq.~\eqref{sigmaBFKL} both impact factors $\imf$ and
eigenvalue function $\chi$ are perturbative objects that admit a
series expansion in $\as$, as in eq.~\eqref{asbchi}; from
next-to-leading order on, they depend on the choice of the energy
scale:
\begin{align}
  \imf^{(j)}(\gamma;p) &= \imf^{(j)}_0(\gamma) + \asb \imf^{(j)}_1(\gamma;p) +
  \ord{\asb^2}\, , \label{PhiBFKL} \\
  \chi(\gamma;p) &= \chi_0(\gamma) + \asb \chi_1(\gamma;p) +
  \ord{\asb^2} \;. \label{chiBFKL}
\end{align}
It turns out that also the impact factors have collinear poles whose
order exceeds that predicted by the collinear dynamics:
$\imf^{(n)}_n(\gamma)\sim\imf^{(j)}_0(\gamma)/\gamma^{2n}$
instead of 
$\imf^{(n)}_n(\gamma)\sim\imf^{(j)}_0(\gamma)/\gamma^{n}$.
Such spurious poles appears when $p\neq 1$, i.e., when the energy
scale is different from $Q_1^2$. However, analogous spurious poles
appear at $\gamma=1$ (anti-collinear limit), unless $s_0=Q_2^2$. The
solution to avoid such spurious poles is to realize that changing
$s_0(p)$ amount to an $\om$-shift in the $\gamma$ variable:
\begin{equation} 
  \bigg(\frac{s}{\kt \kt_0}\bigg)^{\om}  \bigg( \frac{\kt^2}{\kt_0^2}\bigg)^{\gamma}=
  \bigg(\frac{s}{\kt^2}\bigg)^{\om} \bigg( \frac{\kt}{\kt_0}\bigg)^{\om}  \bigg( \frac{\kt^2}{\kt_0^2}\bigg)^{\gamma}=
  \bigg(\frac{s}{ \kt^2}\bigg)^{\om}  \bigg( \frac{\kt^2}{\kt_0^2}\bigg)^{\gamma+\om/2} \;.
  \label{eq:different_scales} 
\end{equation}
This implies that both eigenvalue function and impact factors should
be $\om$ dependent. According to this suggestion, we propose the
following RGI factorization formula:
\begin{subequations}\label{sigmaRGIandGGF}
  \begin{align}
    \sigma^{(jk)}(s,Q_1,Q_2) &= \frac1{2\pi Q_1 Q_2}
    \int\frac{\dif \omega}{2\pi\ui} \left(\frac{s}{s_0(p)}\right)^\om
    \int\frac{\dif \gamma}{2\pi\ui}
    \left(\frac{Q_1^2}{Q_2^2}\right)^{\gamma-\half} \\
    &\hspace{0.2\textwidth} \times
    \Phi^{(j)}(\om,\gamma;p) \,\ggf(\om,\gamma;p)\,\Phi^{(k)}(\om,1-\gamma;-p) \label{sigmaRGI} \\
    \ggf(\om,\gamma;p) &= \frac1{\om-\asb X(\om,\gamma;p)} \;.
    \label{rgiGGF} 
  \end{align}
\end{subequations}

The compatibility of the BFKL factorization formula~\eqref{sigmaBFKL}
with the RGI one~\eqref{sigmaRGI} requires that the one-loop RGI
impact factors at $\om=0$ are related to the BFKL one
\cite{Catani:1994sq,Balitsky:2012bs,Ivanov:2014hpa}
by the relation
\begin{align}  \label{PhiT1om0}
  \Phi_1(0,\gamma) &=
  \frac12\left[\Phi_1(0,\gamma)+\Phi_1(0,1-\gamma)\right] \\
  &= \frac12\left[\imf_1(\gamma)+\imf_1(1-\gamma)
   -\imf_0(\gamma)\de_\om X_0(0,\gamma)
  -\chi_0(\gamma) \big( \de_\om \Phi_0(0,\gamma)
  +\de_\om \Phi_0(0,1-\gamma) \big) \right]\;. \nonumber
\end{align}
We extend $\Phi_1$ at $\om\neq 0$ by requiring the collinear poles to
be located at $\gamma=-\om/2$ and $\gamma=1+\om/2$ and with
$\om$-dependent leading coefficients $M_T(\om)$ and $\bar{M}_T(\om)$ to
be determined by the collinear analysis of the next section, in the form
[$C_0 = \al\as\Big(\sum_q e_q^2\Big) \frac43 T_R\sqrt{2(N_c^2-1)}$]
\begin{equation}\label{phiT1om}
  \Phi_1(\om,\gamma;0) = \Phi_1(0,\gamma)
  +C_0\left[\frac{M_T(\om)}{(\gamma+\hom)^3}
  -\frac{M_T(0)}{\gamma^3}
  +\begin{pmatrix} \gamma\lra 1-\gamma \\ M_T\to\bar{M}_T \end{pmatrix}
  \right] \;.
\end{equation}

\section{Collinear analysis}

\begin{figure}[hbp]
  \centering
  \includegraphics[width=0.25\linewidth]{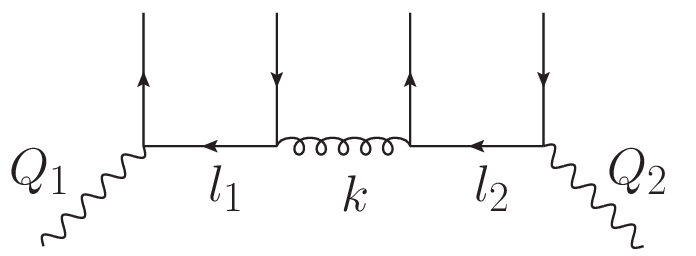}
  \hspace{0.05\linewidth}
\psfrag{(a)}{}
\psfrag{(b)}{}
\raisebox{-2.5ex}{
  \includegraphics[width=0.56\linewidth]{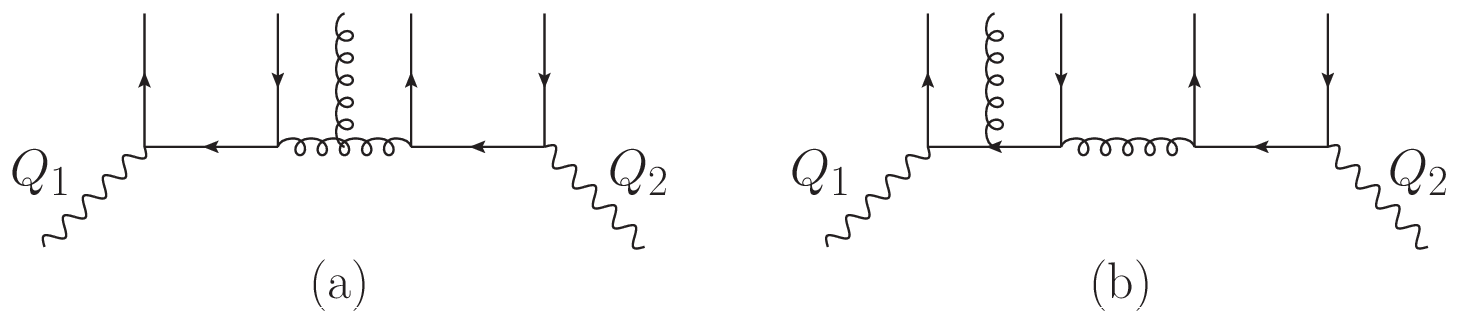}
}
  \caption{Diagramatics of collinear limit at lowest order (left)
    and next-to-leading order (center and right) in the high-energy
    factorization formula.\label{f:colChain}}
\end{figure}

Further information for the $\gamma^*\gamma^*$ cross section, somehow
complementary to the multi-Regge kinematics, can be inferred by analyzing the
collinear limit, i.e., by considering two photons with very different
virtualities, say $Q_1\gg Q_2$. This situation is well described by effective
ladder diagrams, like those depicted in fig.~\ref{f:colChain}, where the
intermediate propagators are strongly ordered in virtuality (decreasing from
left to right). At each QCD vertex, the strong coupling is evaluated at a scale
given by the largest virtuality of the connected propagators, while
the DGLAP~\cite{DGLAP} splitting
functions $P_{ba}(z_b/z_a)$ describes the fragmentation of the parent parton $a$
(to the right) into a child parton $b$ (to the left) and an emitted on-shell parton
(vertical line). The integrals over the ordered longitudinal momentum fractions
are convolutions, which can be diagonalized by a Mellin transform in
the Bjorken variable $1/x_{Bj}=s/Q_1^2=s/s_0(p=1)$.

The cross section in Mellin space $(\om,\gamma)$ stemming from the
left diagram in fig.~\ref{f:colChain} provides the lowest order
integrand of the RGI factorization formula~\eqref{sigmaRGI} in the
collinear limit $\gamma\to0$:
\begin{align}  \label{colChainLO}
  \sigma_0^{(TT)}(\om,\gamma;1)
  &\simeq \Phi^{(T)}_0\,\ggf_0\,\Phi^{(T)}_0\big|_{p=1}\\ 
  &= (2\pi)^3\al \Big(2\sum_{q\in A} e_q^2\Big) \frac1{\gamma}
  \;\cdot\; \frac{\as}{2\pi} \frac{P_{qg}(\om)}{\gamma}
  \;\cdot\; \frac{\as}{2\pi} \frac{P_{gq}(\om)}{\gamma} \;\cdot\;
  \frac{\al}{2\pi} \Big(2\sum_{q\in B} e_{q}^2\Big)
  \frac{P_{q\gamma}(\om)}{\gamma} \;.  \nonumber
\end{align}

Considering ladder diagrams with five splittings between the photons
(last two diagrams of fig.~\ref{f:colChain})
and the running coupling contribution proportional to
$b = \frac{11 N_c-4 T_R N_f}{12\pi}$
provides the integrand of the RGI factorization formula at
$\ord{\as^3}$ in the collinear limit. Altogether, one obtains the
constraint on the coefficients $M_T$ and $\bar{M}_T$ of
eq.~\eqref{phiT1om}:
$M_T(\om)+\bar{M}_T(\om) = 2 P_{qq}(\om)-b$.

Analogous considerations can be done for the longitudinal impact
factor. In this case we obtain
$M_L+\bar{M}_L=\frac{C_F}{T_R} P_{qg}(\om)\frac{3+\om}{2}-b$. In the
last expression, the term proportional to $C_F$ at $\om=0$ is not
consistent with the results of $\phi_1$ of
refs.~\cite{Ivanov:2014hpa}. We don't know the origin of the mismatch,
yet.

\begin{figure}[!htb]
  \centering
  \includegraphics[width=0.45\linewidth,angle=270]{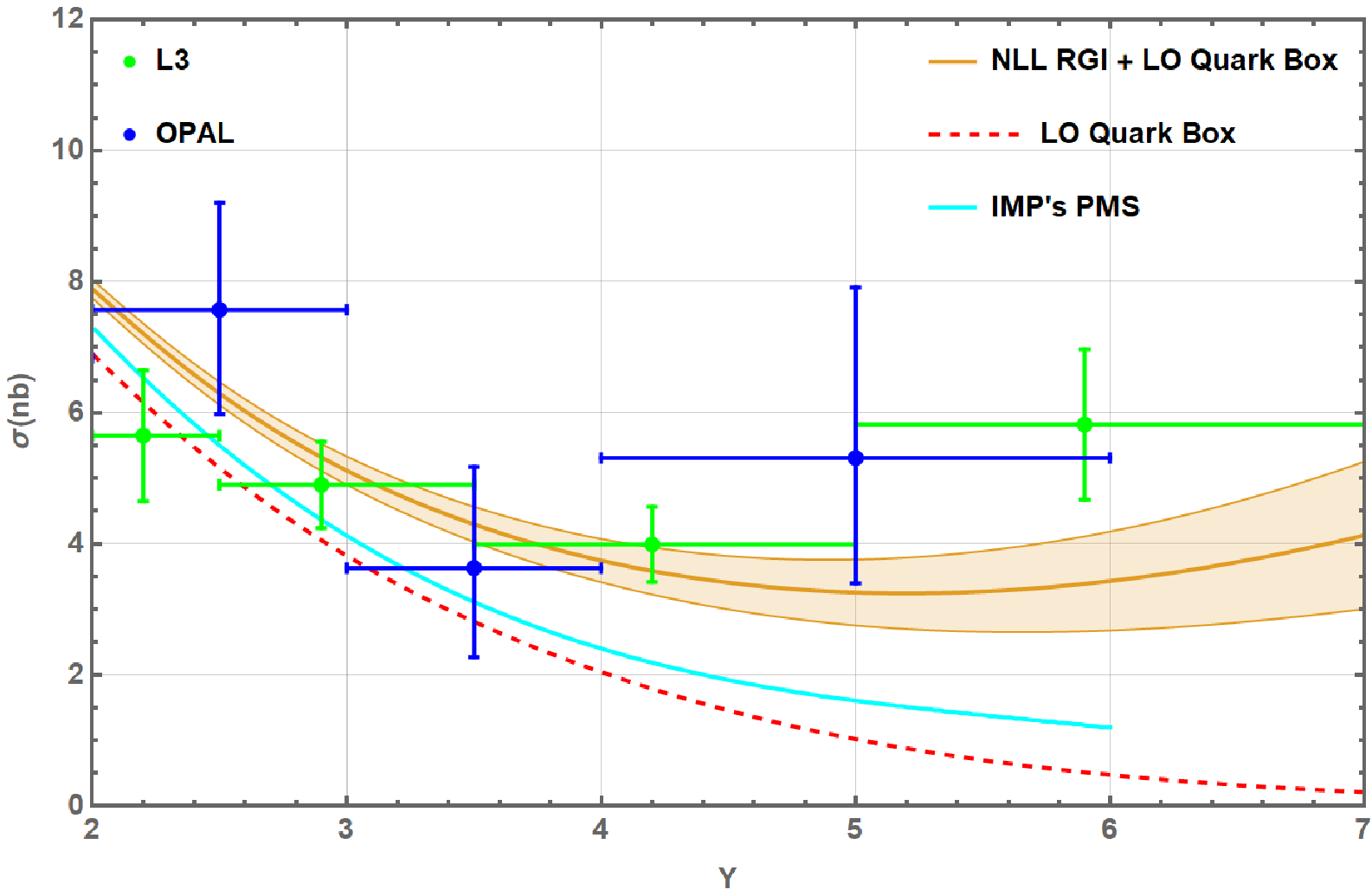}
  \caption{Cross sections for $Q^2 = 17\ \text{GeV}^2$, compared with
    L3 ($Q^2 = 16 \ \text{GeV}^2$) and OPAL ($Q^2 =
    17.9\ \text{GeV}^2$) data. The NLL improved curve is the sum of
    our averaged NLL BFKL resummed scheme and LO quark box
    contribution. The band represents a combination (in quadrature) of
    the scheme uncertainty and the renormalization scale uncertainty,
    The calculation is done for $N_f=4$ massless flavours. The
    Ivanov-Murdaca-Papa's (IMP's) PMS optimized curve (solid-cyan) is
    from \cite{Ivanov:2014hpa}. Separately shown is the quark box contribution
    (dashed red).}
    \label{fig:Q17}
\end{figure}

By evaluating the cross section with the improved impact factor and
GGF for $Q_1^2=Q_2^2=17~\GeV^2$, and by adding the lowest order
perturbative contribution (quark box), we can compare our results with data
measured at LEP~\cite{L3:2001uuv,OPAL:2001fqu}, as can be seen in
fig.~\ref{fig:Q17}.
We also show the LO quark box contribution in this figure.  We observe
that the RGI NLL improved calculation has a
stronger increase over rapidities than the pure NLL one. We also see
that our result is significantly higher than the calculation from
\cite{Ivanov:2014hpa}, particularly at high rapidities.  The RGI
calculation is consistent with the experimental data from LEP within
the theoretical and experimental uncertainties.

\end{document}